\documentstyle[aps,prl,preprint,epsfig]{revtex}    
\tighten

\newcommand{\tgh} {{\rm tgh} \,}

\newcommand{\ket}[1]{\,| \, {#1} \,\rangle  \,}

\newcommand{\be}{\begin{equation}}
\newcommand{\ee}{\end{equation}}
\newcommand{\bq}{\begin{eqnarray}}
\newcommand{\eq}{\end{eqnarray}}

\begin{document}  

\title{Decoherence and $1/f$ noise in Josephson qubits}

\author{E. Paladino$^{(1)}$, L. Faoro$^{(2)}$, 
G. Falci$^{(1)}$, and Rosario Fazio$^{(3)}$}

\address{
$^{(1)}$NEST-INFM $\&$ Dipartimento di Metodologie Fisiche e Chimiche 
	(DMFCI),\\
        Universit\`a di Catania,
        viale A. Doria 6, 95125 Catania, Italy\\
$^{(2)}$ Institute for Scientific Interchange (ISI) $\&$ INFM, \\
	Viale Settimo Severo 65, 10133 Torino, Italy.\\
$^{(3)}$ NEST-INFM $\&$ Scuola Normale Superiore, 
	56126 Pisa, Italy. 
	}

\date{\today}

\maketitle

\begin{abstract}
We propose and study a model of dephasing due to an environment of 
bistable fluctuators. We apply our analysis to   
decoherence of Josephson Qubits, induced by background charges present in 
the substrate, which are also responsible for the $1/f$ noise. 
The discrete nature of the environment leads to a number of new features which
are mostly pronounced for slowly moving charges.
Far away from the degeneracy this model for the dephasing 
is solved exactly.
\end{abstract}

\pacs{PACS numbers: 85.25.Cp, 03.65.Yz, 73.23.-b}

\narrowtext  
Any quantum system during its evolution gets entangled with the surrounding 
environment. This effect is known as {\em decoherence}~\cite{zurek91,palma96}. 
Besides the understanding of its role in many fundamental questions, 
decoherence is studied because it will ultimately limit the performance of a 
quantum computer~\cite{nielsen}.
Solid state nanodevices seem the natural arena to fulfill the 
requirements of large scale integrability and flexibility in the design, 
though, due to the presence of many kinds of low energy excitations in 
the environment, decoherence represents a serious limitation.
Proposals to implement a quantum computer using superconducting 
nanocircuits are proving to be very 
promising~\cite{makhlin99,kn:teor}
and several experiments already highlighted the quantum properties of these 
devices~\cite{kn:exp}.

In superconducting nanocircuits  various sources of decoherence are 
present~\cite{makhlin99,tian00}, as fluctuations 
originating from the surrounding circuit, quasiparticle tunneling, 
fluctuating background charges (BC) and flux noise. 
In this Letter we introduce and study a model for decoherence due to a 
discrete environment, which describes 
what is considered the most serious limitation 
for Josephson qubits in the charge regime, i.e. 
the decoherence originating from fluctuating charged impurities. 
The common belief is that this effect originates from random traps for
single electrons in dielectric materials surrounding the island.
These fluctuations cause the $1/f$ noise directly observed in Single Electron 
Tunneling devices~\cite{zorin96,nakamura-echo}. 

The system under consideration is a  Cooper  pair box~\cite{makhlin99}.
Under appropriate conditions (charging energy $E_C$ much larger than 
the Josephson coupling $E_{J}$ and temperatures $k_B T \ll E_J$ )
only  two charge states are important and the Hamiltonian of the qubit 
${\cal H}_Q$ reads
$$
{\cal H}_Q \,=\, {\delta E_c \over 2} \, \sigma_z  -
 { E_J \over 2} \,\sigma_x \, ,
$$
where the charge basis  $\{\ket{0},\ket{1}\}$ 
is expressed using the Pauli matrices, and 
the bias $\delta E_c \equiv E_{C} (1-C_xV_x/e)$ can be tuned by
varying the applied gate voltage $V_x$.
The environment is a set of BCs electrostatically coupled to the qubit. 
The resulting total Hamiltonian is 
\begin{eqnarray}
	{\cal H}
	&=& {\cal H}_Q \,-\, {1 \over 2} \, \sigma_z \, 
	\sum_i v_i  \,  b_i^{\dagger} b_i \, 
	\, + \, \sum_i 	{\cal H}_i
\label{eq:hamiltonian}\\ 
{\cal H}_i &=& \varepsilon_{c i} b_i^{\dagger} b_i +
		\sum_{k} [T_{ki}  c_{ki}^{\dagger} b_i + \mbox{h.c.} 
]
		 +  \sum_{k} \varepsilon_{ki}  c_{ki}^{\dagger} c_{ki} \, .
\nonumber
\end{eqnarray}
Here ${\cal H}_i$ describes an isolated BC: 
the operators $b_i$ ($b^{\dagger}_i$) destroy (create) an electron 
in the localized  level $\varepsilon_{ci}$.
This electron may tunnel, with amplitude $T_{ki}$ to a band
described by the operators $c_{ki}$, $c^{\dagger}_{ki}$ and the energies 
$\varepsilon_{ki}$. For simplicity  
we assume that each localized level is connected to a distinct band.
An important scale is the total switching rate 
$\displaystyle{\gamma_i=2 \pi {\cal N}(\epsilon_{ci})|T_i|^2}$
(${\cal N}$ is the density of states of the electronic band, 
$|T_{ki}|^2 \approx |T_i|^2$), which characterizes the classical relaxation 
regime of each BC.
Finally the coupling with the qubit is such that each BC  
produces a bistable extra bias $v_i$. The same model for the BCs 
has been used in Ref.~\cite{bauernschmitt93} and
explains experiments on charge trapping in systems of
small tunnel junctions.

Our aim is to investigate the effect of the BC environment on the 
dynamics of the qubit. In this Letter we focus on decoherence {\em during} 
the quantum time evolution, due to an environment which produces $1/f$ noise. 
Thus we are concerned with  BCs with a distribution of 
switching rates $\,\sim 1/\gamma \,$, in the range 
$[\gamma_m,\gamma_M]$~\cite{weissman}. 

We used several techniques as second order perturbation theory in the 
couplings $v_i$, or in the BC-band couplings $T_{ki}$,
Heisenberg equations of motion and a real-time path-integral 
analysis~\cite{kn:cohen-tannoudji,weiss,kn:leggett}.  
The picture which emerges is that the decoherence produced by each BC 
depends qualitatively on the ratio 
$g_i = v_i / \gamma_i$, so it is convenient to distinguish between 
{\em weakly coupled} BCs ($g_i \ll 1$) and 
{\em strongly coupled} BCs (in the other regimes).
We stress that a {\em strongly coupled} BC does not necessarily have large 
coupling $v_i$ with the qubit. 
Indeed in the physical situation we discuss in this Letter 
the energy scale associated to the total extra bias produced by the set 
of BCs is much smaller than the level splitting of the qubit, 
$\sqrt{\delta E_c^2+E_J^2}$. To summarize our results, we find that as far 
as decoherence is concerned, a single weakly coupled BC behaves as a 
source of gaussian noise, whose effect is fully characterized by 
the power spectrum of the unperturbed equilibrium 
fluctuations of the extra bias 
operator $ v_i  \,  b_i^{\dagger} b_i \,$, given by 
$s_i(\omega) =  v_i^2/[2 \cosh^2(\beta \varepsilon_{ci}/2)] \,
\gamma_i /(\gamma_i^2 + \omega^2)$. Each weakly coupled BC contributes 
independently to decoherence. 
Instead our quantum  mechanical treatment  
yields that decoherence due to a strongly coupled BC 
shows pronounced features of its discrete character, as 
saturation effects when $g_i \gg 1$, and dependence on initial conditions. 

We consider, for the sake of clarity, two special operation points for 
the qubit: 
(i) charge degeneracy, $\delta E_c=0$ and 
(ii) the case where tunneling can be neglected, $E_J =0$. 
For this latter case, where pure dephasing occurs without relaxation, 
we find an exact solution.

\begin{figure}
\centerline{{\epsfxsize=9.2cm\epsfysize=6.232cm\epsfbox{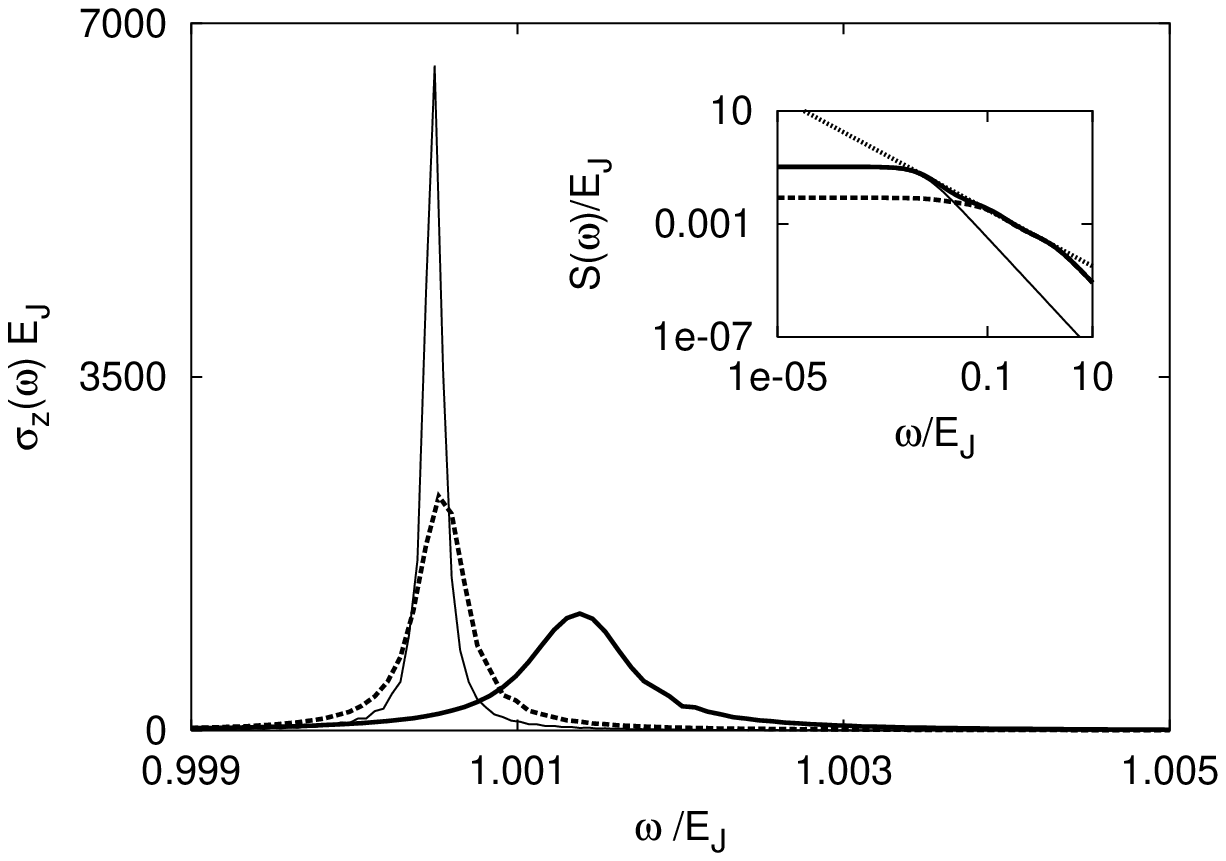}}}
\caption{The Fourier transform $\sigma_z(\omega)$ 
for a set of weakly coupled BCs 
plus a single strongly coupled BC (solid line). The separate effect of 
the coupled BC ($g_0= 8.3$, thin line) and of the 
set of weakly coupled BCs 
(dotted line), is shown for comparison.
Inset: corresponding power spectra.
In all cases the noise level at $E_J$ is fixed to the value 
${\mathrm S}(E_J) / E_J \approx 3.18 \times 10^{-4}$.
}
\label{figure:equil}
\end{figure}

\begin{figure}
\centerline{{\epsfxsize=9.2cm\epsfysize=6.232cm\epsfbox{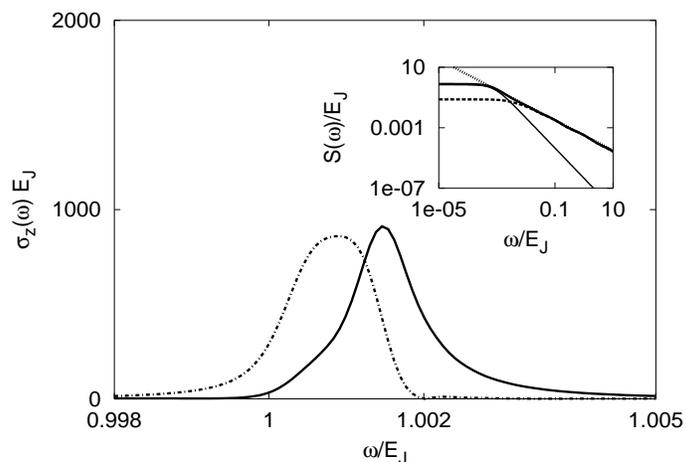}}}
\caption{The Fourier transform $\sigma_z(\omega)$ 
for a set of weakly coupled BCs plus a strongly coupled BC  
($v_0 / \gamma_0 =61.25$) prepared in the ground (dotted line) or in the 
excited state (thick line). 
Inset: corresponding power spectra 
(the thin line corresponds to the extra BC alone).}
\label{figure:nonequil}
\end{figure}

\underline{(i) Charge degeneracy ($\delta E_c=0$) -}  
Relaxation ($\Gamma_r$) and dephasing ($\Gamma_{\phi}$) rates
of the qubit 
given by the Golden Rule~\cite{kn:cohen-tannoudji,weiss}
\begin{eqnarray}
\label{eq:relaxation-dephasing-degeneracy}
\Gamma_{\phi}^{GR} \;=\; \frac{1}{2}\, \Gamma_{r}^{GR} \;=\;
{1 \over 4} \; {\mathrm S} (E_J) 
\end{eqnarray} 
depend only on the power spectrum $S(\omega)= \sum_i s_i(\omega)$ 
at $\omega=E_J$. 
This simple result would readily allow an estimate of 
the rates $\Gamma_{r,\phi}$~\cite{kn:napoli} from independent
measurements of $S(\omega)$~\cite{zorin96,nakamura-echo,kn:noise-set}. 
Being second order in $v_i$, the Golden Rule is appropriate only
for weakly coupled BCs. 
We study the general problem by using the Heisenberg equations of motion. For 
the average values of the qubit observables
$\langle \sigma_{\alpha} \rangle$, (sum over $\alpha,\beta=x,y,z$ is 
implicit) we obtain ($\hbar=1$)
$$
\langle \dot{\sigma}_\alpha \rangle \;=\;
       E_J \;\epsilon_{\alpha \beta} \langle \sigma_\beta \rangle 
       + \eta_{\alpha \beta} \sum_i^N v_i \; \langle b_i^{\dagger} b_i \sigma_\beta \rangle 
$$
where $N$ is the total number of BCs 
and $\epsilon_{yz}= - \epsilon_{zy} = 1$, $\eta_{xy}= - \eta_{yx} = 1$ are the
only nonzero entries.
In the rhs averages of new operators which involve also the 
localized levels and the bands are generated. New equations have to 
be considered and the iteration of this procedure yields an infinite chain. 
A closed set  of $3+3 N$ equations is obtained
by factorization of high order averages:
we ignore the cumulants
$\langle b_i^{\dagger} b_i b_j^{\dagger} b_j \rangle_c$ and 
$\langle b_i^{\dagger} b_i b_j^{\dagger} b_j
\sigma _\alpha \rangle_c$ for $i\neq j$ and insert the relaxation dynamics  
for the BCs in the approximated terms. 
This method gives accurate results for general values of $g_i$ even if 
$v_i / E_J$ is not very small, as we checked by comparing with numerical 
evaluation of the reduced density matrix of the qubit  with few BCs.
Results are presented in Figs.\ref{figure:equil},\ref{figure:nonequil} 
where the time Fourier transform of  $\langle \sigma_z (t)\rangle$, 
proportional to the average charge on the island, is shown. 
We assumed factorized initial conditions for the qubit and the BCs.
We first consider a set of weakly coupled BCs in the range 
$[\, 10^{-2}, 10 \,]\,  E_J$ 
which determine 1/f noise in a frequency interval around the operating 
frequency. 
The coupling strengths $v_i$ have been generated uniformly with approximately 
zero average and magnitudes chosen in order to yield the amplitude of typical 
measured spectra~\cite{zorin96,nakamura-echo,kn:noise-set} 
(extrapolated at GHz frequencies).
These BCs are weakly coupled, and  
determine a dephasing rate which reproduces the prediction
of the Golden Rule Eq.(\ref{eq:relaxation-dephasing-degeneracy}) 
(Fig.\ref{figure:equil} dotted line 
$\Gamma_{\phi}^{GR}/E_J \approx 1.65 \times 10^{-4}$). 
Now we add a slower (and strongly coupled $g_0=v_0/\gamma_0=8.3$) BC, in order
to  extend the 1/f spectrum to lower frequencies.
The added BC gives negligible contribution to the power spectrum at $E_J$
so according to Eq.(\ref{eq:relaxation-dephasing-degeneracy}) it 
should not modify $\Gamma_{\phi}$. Instead, as shown in
Fig.\ref{figure:equil}, we find that the strongly coupled BC alone 
determines a dephasing rate comparable with that of the weakly coupled
BCs. The 
overall $\Gamma_{\phi}$ is more than twice the prediction of the Golden Rule.
If we further slow down the added BC we find that $\Gamma_{\phi}$
increases toward values $\sim \gamma_0$. This indicates that the effect of 
strongly coupled BCs on decoherence saturates (we discuss later similar 
results for the case of pure dephasing, where this conclusion can be made 
sharp).

In this regime we observe also 
effects related to the initial preparation of the strongly coupled BC (see 
Fig.\ref{figure:nonequil}). Finally we checked different sets of BCs
with the same power spectrum. They yield larger decoherence 
if BCs with $g \gtrsim 1$ are present. 
The Golden Rule result 
Eq.(\ref{eq:relaxation-dephasing-degeneracy}) 
{\em underestimates} the effect of these strongly coupled BCs. 
All the features presented above are a direct 
consequence of the discrete nature of the environment. 

\underline{Pure dephasing ($E_J=0$) -}  
In  the absence of 
the tunneling term Eq.(\ref{eq:hamiltonian}) is a model for pure dephasing.
The charge on the qubit
island is conserved,
$[ {\cal H}, \sigma_z ] =0$, but 
superpositions of charge states dephase.
Hence the off-diagonal elements of the reduced density matrix 
of the qubit 
in the charge basis, or equivalently the averages of the 
raising and lowering operators $\sigma_{\pm}$, decay in time. 

It is possible to show by direct calculation that for a product initial 
condition, $\langle \sigma_{\pm} \rangle$ factorizes exactly in independent
contributions of each BC, 
$\langle\sigma_{-}(t)\rangle = \langle\sigma_{-}(0)\rangle 
\prod_j \exp\{-i (\delta E_c - v_j/2)t \} f_j(t)$.
Using a real-time path-integral technique~\cite{weiss}, the general form
of $f_j(t)$ in Laplace space is obtained
\begin{equation}
f_j(\lambda) =  \frac{
\lambda + K_{1,j}(\lambda) 
- i \, v_j / 2 \, \delta p_{j}^0} 
{\lambda^2 + \left ( v_j / 2 \right )^2 
+ \lambda K_{1,j}(\lambda) +
\, v_j / 2\,  K_{2,j}(\lambda)} \, ,
\label{eq:poles-exact}
\end{equation}  
where $\delta p_{j}^0 = 1 -2 \langle b_j^{\dagger}b_j \rangle_{t=0}$.
The kernels $K_{1,j}(\lambda)$ and $ K_{2,j}(\lambda)$
are given by formal series expressions.
An explicit form is obtained in the 
Non Interacting Blip Approximation (NIBA)~\cite{weiss,kn:leggett},   
$K_{1,j}(\lambda)= \gamma_j$, 
$K_{2,j}(\lambda) = - \gamma_j / \pi
[\psi (1/2 +  \beta/2 \pi(\lambda-i \epsilon_{cj})) 
- \psi (1/2 + \beta/2 \pi(\lambda + i \epsilon_{cj}))]$. 
In order to appreciate the validity of the NIBA result, we notice that
it is also obtained by the Heisenberg equations 
with the only assumption that the  
band is in 
equilibrium.

We now discuss the results in the physically relevant limit where
the BCs have an incoherent dynamics. In this case an analytic form for 
$\langle \sigma_{-}(t) \rangle$ is found
\begin{eqnarray} 
\frac{\langle \sigma_{-}(t) \rangle}{\langle \sigma_{-}(0) \rangle}  &=&
\, e^{-i \delta E_c t} \,
\prod_{j=1}^{N}  \, e^{i v_j \, t/2} 
\left \{  A_j \, e^{-{\gamma_j \over 2} (1 - \alpha_j)t}
	\, + \,(1-A_j) \, e^{-{\gamma_j \over 2} (1 + \alpha_j)t} 
	\right \} = \nonumber \\
&& \equiv \exp{ \left \{ -i \delta E_c t \right \}} \, 
\exp{ \left \{ - \Gamma(t) + i E(t) \right \} } \, ,
\label{offdiag}
\end{eqnarray}
where we defined
$
\alpha_j = \sqrt
{1-  g_j^2 - 2i \, g_j	\, 
\tgh \left(\beta \varepsilon_{cj}/2 \right)}  
$ 
and
$
A_j = {1 \over 2 \alpha_j} \, 
\left( 1+ \alpha_j - i \, \delta p_{j}^0 \, g_j \right) \, .
$
This result can be obtained both by approximating the kernel
$K_{2,j}(\lambda) \sim K_{2,j}(0)$, valid in the limit
$\varepsilon_{ci} \, , v_i \, , \gamma_i \ll K_B T $,
and as the exact result of a semiclassical analysis. In this last case
the coupling operator 
$\sum_i v_i b^\dagger_{i} b_{i}$ is substituted  by a
classical stochastic 
process ${\mathcal E}(t)$, sum of 
random telegraph processes, yielding
\begin{equation}
\frac{\langle \sigma_{-}(t) \rangle}{\langle \sigma_{-}(0) \rangle}=
\, e^{-i \delta E_c t} \,
\langle \hskip-2pt \langle \, \exp {[ -i
\int_{0}^{t} \hskip-2pt dt^\prime  {\mathcal E}(t^\prime) ] } \,
\rangle \hskip-2pt \rangle
\label{stocastic}
\end{equation}
where $\langle \hskip-2pt \langle \dots \rangle \hskip-2pt \rangle$ is
the average over the possible realizations of ${\mathcal E}(t)$ 
with given initial conditions $\delta p_j^0$.

The form of Eq.(\ref{offdiag}) 
elucidates the different role of weakly
and strongly coupled BCs in the decoherence process.
Dephasing due to each BC comes from the sum of two exponential terms.
If $g_j \ll 1$ only the first term is important and the 
corresponding rate 
is $\approx 1/[4 \cosh^2(\beta \varepsilon_{cj}/2)] \, v_j^2/\gamma_j$,
the Golden Rule result.
If $g_j \gg 1$  the two terms are of the same order, 
and the decay rate is $\sim \gamma_j$, the switching rate of the BC.
The main effect of strong coupling with the qubit 
is a static energy shift. That is for the slower BCs ($g_j \gg 1$) the 
contribution to the decoherence rate 
saturates to $\sim \gamma_j$.
At short times $t \ll 1/ \gamma_j$ the initial conditions, which 
may take the values $\delta p_j^0 = \pm 1$, determine the transient behavior. 

\begin{figure}
\centerline{{\epsfxsize=9.2cm\epsfysize=6.232cm\epsfbox{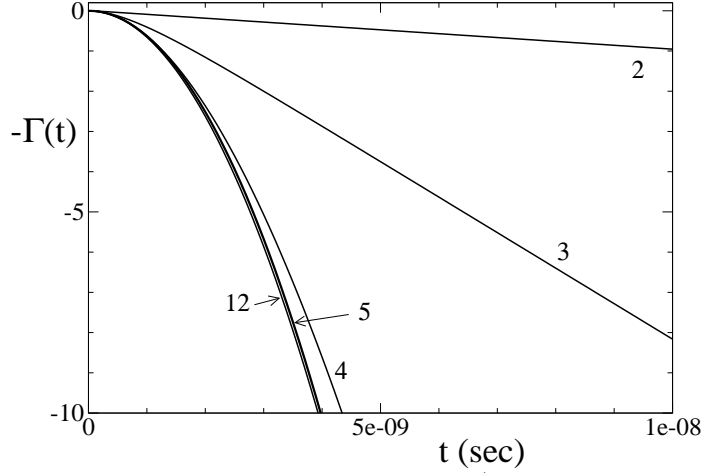}}}
\caption{Saturation effect of slow BCs for a 1/f spectrum.
Relevant parameters ($\langle v \rangle = 9.2 \times 10^7 Hz$, $n_d = 1000$)
give typical experimental measured noise levels and 
reproduce the observed decay of the echo signal~\protect\cite{nakamura-echo}
in charge Josephson qubits.
Couplings $v_i$ are distributed with  dispersion
$\langle \Delta v \rangle / \langle v \rangle = 0.2$. 
$\Gamma(t)$ is almost unaffected by strongly coupled charges 
(the label is the number of decades included). }
\label{dephasing-1overf}
\end{figure}

We now apply this result to sets of BCs which produce $1/f$ noise.  
We stress that while saturation of dephasing due to a {\em single} BC is 
physically intuitive, it 
is not a priori clear whether this holds also for the $1/f$ case,
where a {\em large} number ($\sim 1/\gamma$) of slow fluctuators 
(strongly coupled BCs) is involved.
In Fig.\ref{dephasing-1overf} we 
show the results for a sample with a number of BCs per decade
$n_d=1000$ and with $v_i$ 
distributed with small dispersion around  
$\langle v \rangle = 9.2 \times 10^7 Hz$. Initial conditions 
$\delta p_j^0 = \pm 1$ are distributed according to 
$\langle \delta p_j^0 \rangle = \delta p_{eq}$, the equilibrium value.  
To illustrate the different role played by BCs with 
$g_j \ll 1$ and 
$g_j \gg 1$, we consider  sets with 
$\gamma_M = 10^{12} Hz$ and different $\gamma_m$. 
In this case the dephasing is given by BCs 
with $\gamma_j > 10^7 Hz \approx \langle v \rangle /10$. 
The main contribution comes from three decades at frequencies around 
$\langle v \rangle$. The overall 
effect of the strongly coupled BCs 
($\gamma_j < \langle v \rangle /10$) 
is minimal, despite of their large number. 

\begin{figure}
\centerline{{\epsfxsize=9.2cm\epsfysize=6.232cm\epsfbox{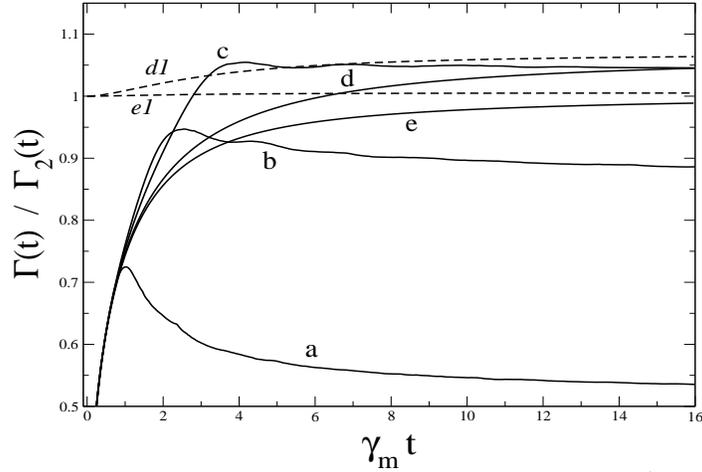}}}
\caption{Ratio 
$\Gamma(t)/ \Gamma_2(t)$ for a $1/f$ spectrum between 
$\gamma_{m}=2 \times 10^7$ and $\gamma_{M}=2 \times 10^9$ with
different numbers of BCs per decade:
(a) $n_d = 10^3$, (b)  $n_d = 4 \times 10^3$,\
(c) $n_d = 8 \times 10^3$, (d) and ({\it d1}) $n_d = 4 \times 10^4$,
(e) and ({\it e1}) $n_d = 4 \times 10^5$.
Full lines corresponds to $\delta p_j^0= \pm 1$, dashed lines to 
equilibrium initial conditions for the BCs.
}
\label{gaussian}
\end{figure}

Finally we compare our results with the gaussian approximation.
It amounts in estimating the average in 
Eq.(\ref{stocastic}) by its second cumulant and taking 
$\delta p_j^0=\delta p_{eq}$~\cite{nota}
\begin{equation}
\Gamma_{2}(t) = \frac{1}{\pi} \int_{0}^{\infty} \hskip-4pt d \omega \;
S(\omega) \;
\frac{1-\cos(\omega t)}{\omega^2} \, .
\label{gauss}
\end{equation}
This formula fails to describe BCs with $g_j \gg 1$. 
For instance
$\Gamma_{2} (t)$ at a fixed $t$, scales with the 
number of  decades and does not show saturation.
The gaussian approximation should become correct if the environment 
has a very large number of extremely weakly coupled BCs. We check 
this limit by comparing $\Gamma_2(t)$  with Eq.(\ref{offdiag}).
The power spectrum $S(\omega)$ is identical for 
all the curves in Fig.\ref{gaussian} but is obtained by sets of charges 
with different 
$n_d$ and $\langle v \rangle$. 
The gaussian behavior is recovered 
for $t \gg 1/\gamma_m$ if  $n_d$ is large (all the BCs are weakly coupled).
If in addition we take $\delta p_j^0 = \delta p_{eq}$ in  Eq.(\ref{offdiag}),
$\Gamma(t)$ approaches $\Gamma_{2} (t)$ also at short
times. 
Hence decoherence depends separately on $n_d$ 
and $\langle v \rangle$, whereas in the gaussian approximation only the 
combination $n_d \langle v^2 \rangle$, which 
enters $S(\omega)$, matters. 
In other words the characterization of the effect of slow sources of $1/f$ 
noise requires knowledge of moments of the bias fluctuations higher than 
$S(\omega)$.

In conclusion we studied decoherence due to an environment of bistable charges. 
We found that the average coupling 
between individual BCs and the qubit is an important 
scale of the problem: BCs such that $\gamma_i \ll v_i$ show 
pronounced features of their discrete dynamics, as saturation 
and transient behavior. 
The physical picture we obtain for the decoherence effects due to BCs
is not sensitive to the details of the model Hamiltonian 
(\ref{eq:hamiltonian}), but is mainly determined by the
discrete character of the environment.  
Thus this approach can also be applied to different physical systems
as phase qubits in the presence of flux noise~\cite{tian00}.

We thank G. Castelli, Y. Nakamura,
M. Palma, M. Rasetti, A. Shnirman, M. Sassetti, D. Vion, 
U. Weiss and A. Zorin for discussions.
This work has been supported by EU (IST-FET SQUBIT) and INFM-PRA-SSQI.

\begin{thebibliography}{299}
\bibitem{nota}
Eq.(\ref{gauss}) can be also obtained by replacing the BCs with 
a set of harmonic oscillators 
with the same $1/f$ spectrum. See ref.~\cite{palma96} and W. Unruh, Phys. Rev. A, {\bf 51},
992 (1995).
\end{thebibliography}
\end{document}